\documentstyle[11pt]{article}
\title{Correlation effects in MgO and CaO: \\Cohesive energies and lattice
constants}

\author{Klaus Doll, Michael Dolg \\ \\
Max-Planck-Institut f\"ur Physik komplexer Systeme \\
D-01187 Dresden, Germany \\ \\  Hermann Stoll
\\ \\ Institut f\"ur Theoretische Chemie \\
        Universit\"at  Stuttgart\\
        D-70550 Stuttgart,  Germany \\}
\begin{document}
\begin{titlepage}
\maketitle
\begin{abstract}
\noindent
A recently proposed computational scheme based on local increments has been applied to the calculation of correlation
contributions to the cohesive energy of the CaO crystal. Using {\em ab-initio} 
quantum chemical methods for evaluating individual increments, we obtain $\sim 80\%$ of the
difference between the experimental
and Hartree-Fock cohesive energies. Lattice constants corrected for correlation effects deviate by less than
1\% from
experimental values, in the case of MgO and CaO.
\end{abstract}
accepted by Phys. Rev. B
\end{titlepage}

\section{Introduction}

{\em Ab-initio} Hartree-Fock (HF) and configuration interaction (CI) methods are standard tools in computational
chemistry nowadays and various program packages are available for accurate 
calculations of properties of atoms and molecules. For solids,
HF calculations have become possible, on a broad scale, with the advent of the program
package CRYSTAL \cite{CRYSTAL}. However, the problem of an accurate treatment of electron
correlation  is not fully settled
(for a survey see \cite{FuldeBuch}).

Although the absolute value of the HF energy is usually much larger than 
the correlation energy, the correlation energy is very important for
energy differences. For example, the O$^-$ ion is not stable at the HF level,
and correlations are necessary in order to obtain even qualitative agreement with the 
experimental result for the electron affinity of oxygen. In solid state physics, NiO is a well known example
of a system which is insulating due to correlations.

The most widely used method to include correlations
in solids is density functional theory (DFT) \cite{DFT}.
DFT has also recently become quite popular for a computationally efficient
treatment of exchange and correlation in molecules.
However, a systematic improvement towards the exact results is currently not possible with DFT.
Wave-function-based methods are more suitable for this purpose.

In the last years,
Quantum Monte-Carlo  
calculations have been performed for several
systems \cite{QMC}. 
Correlations are included here by multiplying the HF
wavefunction with a Jastrow factor. 
An approach more closely related
to quantum chemistry is the Local Ansatz \cite{Stollhoff,FuldeBuch} 
where judiciously chosen local excitation operators are applied to
HF wavefunctions from CRYSTAL calculations. 
Some years ago, an incremental scheme has been
proposed and applied in calculations for semiconductors 
\cite{StollDiamant,Beate}, for graphite \cite{StollGraphit} and for
the valence band of diamond \cite{Juergen}; here information on the effect of local excitations on solid-state properties is drawn
from calculations using
standard quantum chemical program packages. In a recent paper \cite{DDFS}
we showed that this method can be successfully extended to ionic solids;
we reported results for the correlation contribution to the cohesive energy of MgO.
In the present article,
we apply the scheme to the 
cohesive energy of CaO, as a second example. In addition, we show
how correlations affect the lattice constants of
MgO and CaO.
For these systems, several calculations have been performed at the HF level 
with 
the CRYSTAL code \cite{CausaMgO1,DRFASH,MackrodtCaO,CausaZupan,McCarthy,Catti}
as well as with inclusion of correlations using DFT
\cite{DRFASH,CausaZupan,McCarthy,Pyper}.

\section{The Method}

The method of increments can be used to build up correlation effects in solids
from local correlation contributions which in term may be obtained by
transferring results from suitably embedded finite clusters to the infinite crystal.
It has been fully described in \cite{StollDiamant,Beate,StollGraphit,DDFS}, and a formal derivation has been given
within the framework of the projection technique \cite{TomSchork}. Thus,
we will only briefly repeat the main
ideas.\\
(a) Starting from  self-consistent field (SCF)
 calculations localized orbitals are
generated which are assumed to be similar in the clusters and in the solid. \\
(b) One-body correlation-energy increments are calculated: in our specific case
these are the correlation energies $\epsilon (A)$, 
$\epsilon (B)$, $\epsilon (C)$, ... of localized orbital groups which can be attributed to X$^{2+}$ (X = Mg, Ca) or O$^{2-}$ ions
at ionic positions $A$, $B$, $C$, ...   Each localized orbital group is correlated separately. \\
(c) Two-body increments are defined as non-additivity corrections: 
\begin{eqnarray*}
\Delta \epsilon {(AB)}=\epsilon (AB) - \epsilon (A) -\epsilon (B),
\end{eqnarray*}
where $\epsilon {(AB)}$ is the correlation energy of the joint
orbital system of $AB$.\\
(d) Three-body increments are defined as
\begin{eqnarray*}
\Delta\epsilon(ABC) = & \epsilon(ABC) -   
[\epsilon(A) + \epsilon(B) + \epsilon(C)] - \nonumber  \\
  &  [\Delta\epsilon(AB) + \Delta\epsilon(AC) + \Delta\epsilon(BC)].
\end{eqnarray*}
Similar definitions apply to higher-body increments. \\
(e) The correlation 
energy of the solid can now be expressed as the sum of all possible increments:
\begin{equation}
\epsilon_{\rm bulk} = \sum_A \epsilon(A) + \frac{1}{2} \sum_{A,B} \Delta\epsilon(AB) + \frac{1}{3!} \sum_{A,B,C}
\Delta\epsilon(ABC) + ...
\end{equation}
Of course, this 
only makes sense if the incremental expansion is well
convergent, i.e. if $\Delta \epsilon (AB)$ rapidly decreases with
increasing distance of the ions at position $A$ and $B$ and if the
three-body terms are significantly smaller than the two-body ones.
A pre-requisite is that the correlation method used for evaluating the increments must be size-extensive:
otherwise the two-body increment $\Delta \epsilon (AB)$
for two ions $A$ and $B$ at infinite distance would not vanish. 
In our present work, we used three different size-extensive approaches, cf.
 Sect. 2.1.
Finally, the increments must be transferable, i.e.\ they should only weakly depend
on the cluster chosen. 
\subsection{Correlation Methods}
In this section we want to give a brief description of the correlation
methods used.
In the averaged coupled-pair functional (ACPF \cite{ACPF}) 
scheme, the correlation energy is expressed in the form
\begin{equation}
E_{\rm corr}[\Psi_c]=\frac{< \Psi_{SCF}+\Psi_c|H-E_{SCF} |\Psi_{SCF}+\Psi_c >}
{1+g_c<\Psi_c|\Psi_c>}
\end{equation}
with $\Psi_{SCF}$ being the SCF-wavefunction
(usually of the spin-restricted Hartree-Fock type)
 and $\Psi_c$ the correlation part of the 
wavefunction, 
\begin{eqnarray}
|\Psi_c>= \sum_{a \atop r}c_{a}^{r}a^+_r  a_a|\Psi_{SCF}> +
\sum_{a<b \atop r<s}c_{ab}^{rs}a^+_r a^+_s a_a a_b|\Psi_{SCF}>;
\end{eqnarray}
$g_c$ is chosen as $\frac{2}{n}$ in order to make the expression (2) 
approximately size-consistent
($n$ being the number of correlated electrons).
For more details (and the extension to multi-reference cases), see \cite{ACPF}.\\
In the coupled-cluster singles and doubles (CCSD \cite{CCSD}) scheme, 
the wavefunction is expressed with the help of an
exponential ansatz:
\begin{eqnarray}
|\Psi_{CCSD}>=\mbox{exp}({\sum_{a \atop r}c_a^r a^+_r a_a +
\sum_{a<b \atop r<s}c_{ab}^{rs}a^+_r a^+_s a_a a_b})|\Psi_{SCF}>.
\end{eqnarray}
$a^+$ ($a$) are creation (annihilation) operators of electrons in 
orbitals which are occupied ($a$, $b$) or unoccupied ($r,$ $s$) in
the SCF wavefunction.\\
Finally, in the CCSD(T) scheme, three particle excitations are included 
by means of perturbation theory as proposed in \cite{Raghavachari}.\\
We used these three methods to compare their quality
in applications to solids. It turns out that 
ACPF and CCSD give very similar results, while CCSD(T) yields slighty improved energies
\cite{CCSDTWerner}.
Altogether, the results are not strongly dependent on the methods and no
problem arises, therefore, if only one method should be applicable (as is the case
for low-spin open-shell systems, where CCSD and CCSD(T) are not yet 
readily available).
All calculations of this work were done by using the program package MOLPRO 
\cite{KnowlesWerner,MOLPROpapers}.

\section{Cohesive Energy of CaO}

\subsection{Basis sets and test calculations}

For calculating the correlation
contribution to the cohesive energy of CaO, we closely follow the approach of Ref.\ \cite{DDFS}. For oxygen we choose
a $[5s4p3d2f]$ basis set \cite{Dunning}. 
Calcium is described by a small-core pseudopotential replacing the
$1s$, $2s$ and $2p$ electrons \cite{KauppCa},  
and a corresponding $[6s6p5d2f1g]$ valence basis set
(from ref. \cite{KauppCa}, augmented with polarization functions  $f_1$=0.863492, $f_2$=2.142 and
$g$=1.66) is used. 
All orbitals are correlated, with the exception of the O $1s$ core.
In particular, the correlation contribution of the outer-core Ca $3s$ and $3p$ orbitals
is explicitly taken into account.
We did not use a large-core (X$^{2+}$) pseudopotential and a core
polarization potential (CPP) for treating core-valence correlation as was done in the
case of  
MgO \cite{DDFS}, because Ca is close to the transition metals and 
excitations into  
$d$-orbitals are important. The influence of the latter on the X$^{2+}$ core cannot be well represented
by a CPP since the $3d$ orbitals are core-like themselves (cf.\ the discussion in
\cite{MuellerFleschMeyer}). Correlating the Ca outer-core orbitals explicitly, using the small-core (Ca$^{10+}$) pseudopotential,
we circumvent this problem.

Using this approach, we performed test calculations for the
first and second ionization potential of the Ca atom (Table 1) and calculated
spectroscopic properties of the CaO molecule (Table 2).
In both cases, we obtain good agreement with experiment.

\subsection{Intra-atomic Correlation}
We first calculated one-body correlation-energy increments. For Ca$^{2+}$,
the results are virtually independent of the solid-state surroundings. This was tested by
doing calculations for a free $\rm {Ca}^{2+}$ and a $\rm {Ca}^{2+}$ 
embedded in point charges. (A cube of 7x7x7 ions was simulated by point 
charges $\pm$2, with charges
at the surface planes/edges/corners reduced by factors 2/4/8, respectively.)

In the case of $\rm {O}^{2-}$, of course, the solid-state influence is decisive for stability, and we took it
into account by using an embedding similar to that of Ref.\
\cite{DDFS}: the Pauli repulsion of the 6 nearest $\rm {Ca}^{2+}$ neighbours was simulated by
large-core pseudopotentials \cite{Fuentealba}, while the rest
of a cube of 7x7x7 lattice sites was treated in point-charge approximation again. 
A NaCl-like structure with a lattice constant of 4.81 \AA \mbox{ } was adopted.
(The experimental value for the
lattice constant is 4.8032 \AA \mbox{ } at a temperature of T=17.9 K 
\cite{LandoltGitter}).
We performed similar calculations for various other finite-cluster approximations of the CaO crystal, in order
to insure that the results are not sensitive to lattice extensions beyond the cube mentioned above.

The results for the one-body correlation-energy increments are shown in Table 3.
It is interesting to note that the absolute value of the Ca $\rightarrow 
\rm{Ca}^{2+}$
increment is larger than in the case
of Mg, although the electron density in the valence region of Ca is lower
than for Mg. 
The larger correlation contribution for Ca can be rationalized by the fact that excitations into
low-lying unoccupied $d$-orbitals are much more important for Ca than for Mg.
This is a result which would be difficult to explain by
density functional theory: in a local-density framework, higher density leads to a higher
absolute value of the 
correlation energy. 

In Ref.\ \cite{DDFS} we argued that the increment  in correlation energy 
$\epsilon(\rm {embedded}$ $\rm O^{2-})-\epsilon(\rm {free}$ ${\rm O})$ is not 
just twice the
increment $\epsilon(\rm {free}$ $\rm O^{-})-\epsilon(\rm {free}$ $\rm O)$.
However, comparing the increments 
$\epsilon$ (embedded $\rm {O}^{2-})-\epsilon (\rm {embedded} $ $\rm O)$
and $\epsilon$ ( {embedded} $\rm {O}^{-})-\epsilon (\rm embedded $ ${\rm O})$
one finds a factor very close to two.
This can be seen from Table
4 where we compare the increments in the case of MgO.
Thus, for the embedded species linear scaling is appropriate as in the case of the gas-phase  isoelectronic series
Ne, Ne$^+$, Ne$^{2+}$: there, the increments in correlation energy are
0.0608 H (Ne$^{2+}$ $\rightarrow$ Ne$^+$) and 0.0652 H (Ne$^{+}$ $\rightarrow$ Ne) 
\cite{Davidson}.
Table 4 also shows that the correlation contribution to the electron affinity of the oxygen atom
is {\em smaller} for the embedded species than in the gas phase.
This is due to the fact that energy differences to excited-state configurations become larger
when enclosing O$^{n-}$ in a solid-state cage.
Once again, this is at variance with a LDA description
as the electron density in the  case
of the embedded $\rm O^-$ 
is more compressed than in the case of a free $\rm O^-$.

In Figure 1 we show the charge density distribution of
O$^{2-}$, again in the case of MgO. We used basis functions on both O and Mg; 
the Mg $1s$, $2s$ and $2p$-electrons are replaced by a pseudopotential.
One recognizes the minimum near the Mg$^{2+}$ cores, where the 
Pauli repulsion prevents the oxygen electrons from penetrating into the Mg$^{2+}$
core region. This way, the solid is stabilized.
The 6$^{th}$ contour line, counting from Mg to O, is the line which 
represents a density of 0.002 a.u. This is the density which 
encloses about 95 \% of the charge and was proposed as an estimate
of the size of atoms and molecules \cite{Bader}.

The sum of the intra-ionic correlation-energy increments discussed in this subsection
turns out to yield only $\sim$ 60 \% of
the correlation contribution of the cohesive energy of CaO. This percentage is quite similar to that
obtained for MgO  \cite{DDFS}, at the same level. 
Thus, although MgO and CaO are to a very good approximation purely ionic solids,
the inter-atomic correlation effects to be dealt with in the next subsection play an important role.

\subsection{Two- and three-body increments}
When calculating two-body 
correlation-energy increments, point charges or pseudopotentials surrounding 
a given ion have to be replaced by 
'real' ions. In the case of an additional 'real' $\rm {O}^{2-}$, its next-neighbour shell also
has to be replaced by a cage of pseudopotentials simulating $\rm Ca^{2+}$. 
This way the increments shown in Table 3 are obtained. 

It turns out that the Ca-O increments are much more important than 
the O-O increments, while Ca-Ca increments are negligibly small. The changes with respect to
MgO \cite{DDFS} can easily be rationalized: On the one hand, the
lattice constant is larger than in the case of MgO (4.81 \AA \mbox{ }
vs. 4.21 \AA),
which reduces the van der Waals interaction and makes the O-O increments
smaller. On the other hand, the polarizability of $\rm Ca^{2+}$ is by a 
factor of more than 6
higher than that of $\rm Mg^{2+}$
(see for example Ref. \cite{Fuentealba}), 
which leads to large Ca-O increments. 
We show the van der Waals-like decay in Figure 2
by plotting the two-body increments
O-O for CaO from CCSD calculations (without including weight
factors). By multiplying with the sixth power of the distance, one can
verify the van der Waals-law. Plots for the other two-body increments
are qualitatively similar.

Three-body increments contribute with less
than 2 \% to the correlation piece of the bulk cohesive energy and may safely be neglected, therefore.
A survey of the convergency pattern of the incremental expansion, for both CaO and MgO, is given in Figs.\ 3 and 4.

\subsection{Sum of increments}

Adding up the increments of sections 3.2 and 3.3
(cf.\ Table 5), we obtain between 71 and 78 \% of the 
'experimental' correlation contribution to the cohesive energy 
which we define as the difference of the experimental cohesive energy (11.0 eV,
\cite{CRC})
plus the zero-point energy (which is
taken into account within the Debye approximation
and is of the order 0.1 eV) minus
the HF binding energy  (7.6 eV, Ref. \cite{MackrodtCaO}). The percentage 
obtained is slightly
less compared to the case of MgO \cite{DDFS} where 79 to 86 \% were
recovered. One of the reasons for this difference is
that we used a CPP in the case of MgO which covers nearly 100\% of the 
core-valence correlation contributions in Mg, while the
explicit treatment of that correlation piece for Ca was less exhaustive.
Another reason is that on the
Hartree-Fock level $f$-functions for Ca (which are not yet implemented in
CRYSTAL) would probably increase the cohesive energy and lower the 
'experimental' correlation contribution.
Finally, as in the case of MgO, a significant part of the missing correlation 
energy should
be due to basis set errors for the O atom. --
The total cohesive energy 
recovered in our calculations is in the range of between 91 and 93 \% of the 
experimental value.

Our results are compared in Table 5
to those from density functional calculations. We
choose the results from \cite{DRFASH} where a correlation-only functional
was used and  0.078 to 0.097 H  of the correlation contribution to the 
cohesive energy were obtained, depending on the specific correlation functional used.

\section{Lattice constants}

At the Hartree-Fock level, the lattice constant is in good agreement with
experiment for MgO
\cite{CausaMgO1,DRFASH,CausaZupan,McCarthy,Catti}, 
whereas there is a deviation of 0.05 \AA \mbox{ }
in the case of CaO \cite{MackrodtCaO}. 
It is interesting, therefore, to study the influence of correlation effects on lattice
constants. In Tables 6 and 7, we give the necessary increments
for MgO and CaO, respectively.
We find two
main effects of correlations. On the one hand,  the van der 
Waals-interaction leads to a reduction of the lattice spacing since the
attractive interaction is of the form $-\frac{1}{r^6}$ 
and obviously stronger at shorter distance. On the other hand, we find
that the intra-ionic correlation of the $\rm O^{2-}$-ion forces a larger constant. This can be 
understood from the argument that excited configurations are lower in energy
and mix more strongly  with the ground-state
determinant if the
$\rm O^{2-}$ is less compressed as explained in section 3.2. 

Adding up all these contributions (cf.\ Table 8), they are found to nearly cancel
in the case of MgO and to lead to a reduction of only 0.01 \AA \mbox{ }.
For obtaining this result, we applied a linear fit to the correlation energy
and superimposed it on the HF potential curve of Refs. \cite{Catti,MackrodtCaO}.
We checked the validity of the linear approximation by calculating
selected increments at other lattice constants.

In the case of CaO, the van der Waals-interaction is more important and
the lattice constant is reduced to 4.81 \AA \mbox{ } which is in nice
agreement with the experimental value.
The lattice constants seem to be in better agreement with the experimental
values than those calculated from density functional theory  for MgO
\cite{CausaZupan,McCarthy}  and CaO \cite{CausaZupan}, where deviations 
of $\pm$ 2 \% are found. This is similar to earlier findings for
semiconductors \cite{Beate}.

\section{Conclusion}

We determined the correlation contribution to the cohesive energy of CaO
using an expansion into local increments
recently applied to MgO. Making use of quantum-chemical {\em ab-initio} 
configuration-interaction
calculations for evaluating individual increments, 
we obtain $\sim$ 80 \% of the 
expected value. The missing energy is probably mainly due to 
the lack of $g$ and higher polarization functions in our one-particle basis set.
The computed lattice constants show  deviations of less than 1\%
from the experimental values. We found two correlation effects on the lattice constants:
the inter-atomic van der Waals-force leads to a reduction, whereas intra-atomic
correlations of the $\rm O^{2-}$ ions lead to an increase of the lattice constant.

The main difference between CaO and MgO is the reduced importance of
the inter-atomic O-O correlations in CaO (due to the larger lattice 
constant) and the higher importance of the Ca-O correlations (due to
the higher polarizability of Ca$^{2+}$).

Compared to DFT, the numerical effort of our scheme is significantly higher. However, we feel 
that the advantage of the present approach is the high quality and stability
of the results both for atoms, ions as well as for solids. Another advantage
is the possibility of a systematic improvement by using larger basis sets.

We think that the method of local increments is capable now of being routinely applied to ionic
systems, and a systematic study on alkali halides is underway. An 
extension to open-shell systems such as NiO is also a project currently under
investigation.

\section*{Acknowledgments}
We would like to thank Prof.\  P.\ Fulde for supporting this work and
Prof.\ W.\ C.\ Nieuwpoort (Groningen)
for interesting suggestions.
We are grateful to Prof.\ H.-J.\ Werner (Stuttgart) for providing the 
program package MOLPRO.

\newpage
\begin{table}
\begin{minipage}{5in}
\begin{center}
\caption{\label{2}Atomic ionization potentials Ca$\rightarrow$Ca$^+$/Ca$^+$$\rightarrow$Ca$^{2+}$ (in eV)}
\vspace{5mm}
\begin{tabular}{|c|c|}
\hline
\hline
RHF
& 5.16/11.35 \\
ACPF & 6.01/11.78  \\
CCSD & 6.03/11.77  \\
CCSD(T) & 6.09/11.80 \\ \hline
expt. \cite{Moore} & \multicolumn{1}{|c|}{6.11/11.87} \\
\hline\hline
\end{tabular}
\end{center}
\end{minipage}
\end{table}

\begin{table}
\begin{center}
\begin{minipage}{5in}
\caption{\label{3}Bond length $R_e$ (\AA), dissociation energy $D_e$ (eV) and vibrational frequency
$\omega_e$ (cm$^{-1}$) of the CaO
molecule.} 
\vspace{5mm}
\begin{tabular}{|c|ccc|}
\hline\hline
 &  $R_e$ & \multicolumn{1}{c}{$D_e$} & $\omega_e$ \\
\hline
RHF
& 1.812 & 0.67 & 829 \\
CCSD & 1.822 & 3.39 & 769  \\
CCSD(T) & 1.846  & 3.84  & 681 \\
\hline
expt. \cite{Huber-Herzberg,Irvin} & 1.822 & 4.16$\pm$0.07 & 732.1 \\ 
\hline\hline
\end{tabular}
\end{minipage}
\end{center}
\end{table}

\newpage
\begin{table}
\begin{center}
\begin{minipage}{5in}
\caption{\label{5}Local increments (a.u.) for CaO at a lattice constant
of 4.81 \AA.}

\vspace{5mm}
\begin{tabular}{|c|c|rrr|}
\hline\hline
 & \multicolumn{1}{|c|}{weight\footnote{Weight factor in the incremental expansion of the bulk correlation
  energy (in a.u. per primitive unit cell) of CaO.}} 
 & \multicolumn{1}{|c}{ACPF
 } & \multicolumn{1}{c}{CCSD} & \multicolumn{1}{c|}{CCSD(T)}\\
\hline\hline
$\rm {Ca} \rightarrow \rm {Ca^{2+}}$ & 1 & +0.047045  &  +0.047359 &  
+0.050921 \\
\hline
$\rm {O}  \rightarrow \rm {O^{2-}}$ & 1 & -0.096104 & -0.097083 & -0.102340 \\
\hline\hline
\multicolumn{2}{|c|} {sum of one-body increments}   
& -0.049059 & -0.049724 & -0.051419 \\
\hline\hline
$\rm{Ca-O}$ next neighbour & 6 & -0.037704 & -0.035436 & -0.040266 \\
\hline
$\rm{Ca-O}$, $2^{nd}$ next neighbour & 8 & -0.000880 & -0.000928 & -0.001056 \\
\hline
$\rm{Ca-O}$, $3^{rd}$ next neighbour & 24 & -0.000288 & -0.000504 & -0.000576 \\
\hline
$\rm{Ca-O}$, $4^{th}$ next neighbour & 30 & -0.000150 & -0.000090 & -0.000120 \\
\hline
$\rm{Ca-Ca}$, next neighbour & 6 & -0.001002 & -0.001026 & -0.001128 \\
\hline
$\rm{Ca-Ca}$, $2^{nd}$ next neighbour & 3  & -0.000054 & -0.000057 & -0.000063 \\
\hline
$\rm{Ca-Ca}$, $3^{rd}$ next neighbour & 12 & -0.000060 & -0.000072 &
-0.000072 \\ \hline
$\rm{O-O}$, next neighbour & 6 & -0.006888 & -0.006402 & -0.007548 \\
\hline
$\rm{O-O}$, $2^{nd}$ next neighbour & 3 & -0.000363 & -0.000345 & -0.000402 \\
\hline
$\rm{O-O}$, $3^{rd}$ next neighbour & 12 & -0.000324 & -0.000312 & -0.000360 \\
\hline
$\rm{O-O}$, $4^{th}$ next neighbour & 6 & -0.000066 & -0.000066 & -0.000078 \\
\hline
$\rm{O-O}$, $5^{th}$ next neighbour & 12 & -0.000060 & -0.000060 & -0.000078 \\
\hline
$\rm{O-O}$, $6^{th}$ next neighbour & 4 & -0.000016 & -0.000008 & -0.000016 \\
\hline
\hline
\multicolumn{2}{|c|} {sum of two-body-increments} & -0.047855  & -0.045306 & -0.051763 \\
\hline\hline
$\rm {O-O-O}$ \footnote{ions at (1,0,0), (0,1,0) and (0,0,1)}
 & 8 & +0.000224 & +0.000176 & +0.000200 \\
\hline
$\rm {O-O-O}$ \footnote{ions at (1,0,0), (-1,0,0) and (0,0,1)}
 & 12 & +0.000060 & +0.000036 & +0.000048 \\
\hline
$\rm {O-Ca-Ca}$ \footnote{O at (0,0,0), Ca at (0,0,1) and (0,1,0)}
 & 12 & +0.000036 & +0.000288 & +0.000276 \\
\hline
$\rm {O-Ca-Ca}$ \footnote{O at (0,0,0), Ca at (0,0,1) and (0,0,-1)}
 & 3 & -0.000078 & -0.000015 & -0.000030 \\
\hline
$\rm {O-Ca-O}$ \footnote{O at (0,0,0) and (0,1,1), Ca at (0,1,0)}
  & 12 & +0.000792 & +0.001092 & +0.000972 \\
\hline
$\rm {O-Ca-O}$ \footnote{O at (1,0,0) and (-1,0,0), Ca at (0,0,0)}
 & 3 & -0.000180 & -0.000060 &
-0.000132 \\
\hline
$\rm {O-Ca-O}$ \footnote{O at (0,0,0) and (0,1,1), Ca at (1,0,0)}
 & 24 & -0.000048 & +0.000216 &
+0.000192 \\
\hline\hline
\multicolumn{2}{|c|} {sum of three-body increments} & +0.000692  & +0.001685 &  +0.001412 \\
\hline\hline
\multicolumn{2}{|c|} {total sum} & -0.096342 & -0.093445 & -0.101910 \\
\hline\hline
\end{tabular}
\end{minipage}
\end{center}
\end{table}

\newpage
\begin{table}
\begin{minipage}{5in}
\begin{center}
\caption{\label{1}Intra-ionic correlation of free and embedded oxygen (in a.u.). }
\begin{tabular}{|c|c|c|}
\hline\hline
 & incr. O $\rightarrow$ $\rm {O}^-$ & incr. O $\rightarrow$ $\rm {O}^{2-}$\\
\hline
O and $\rm {O}^-$ free, $\rm {O}^{2-}$ embedded & 0.062794 &  0.096460 \\
O, $\rm {O}^-$, $\rm {O}^{2-}$ embedded & 0.050474 & 0.100715 \\
\hline\hline
\end{tabular}
\end{center}
\end{minipage}
\end{table}

\begin{table}
\begin{center}
\begin{minipage}{5in}
\caption{\label{66} Correlation contributions to the cohesive energy  of CaO (in 
a.u.).}
\vspace{5mm}
\begin{tabular}{|ccc|c|c|}
\hline\hline
 ACPF  & CCSD & CCSD(T) & DFT & expt.  \\ \hline
 0.095 & 0.092 & 0.101 & 0.078 ... 0.097 \cite{DRFASH} & 0.129 \\ \hline
\hline
\end{tabular}
\end{minipage}
\end{center}
\end{table}

\newpage
\begin{table}
\begin{center}
\begin{minipage}{5in}
\caption{\label{77}Local increments (in a.u.) for MgO
 at a lattice constant of 4.18 \AA }
\vspace{5mm}
\begin{tabular}{|c|rrr|}
\hline\hline
 & \multicolumn{1}{|c}{ACPF} & 
 \multicolumn{1}{c}{CCSD} & \multicolumn{1}{c|}{CCSD(T)}\\
\hline\hline
$\rm {Mg} \rightarrow \rm {Mg}^{2+}$ & +0.046897  &  +0.046897  & +0.046897 \\
$\rm {O}  \rightarrow \rm {O^{2-}}$ & -0.094833 & -0.095808 & -0.100867 \\
\hline
\multicolumn{1}{|c|} {one-body increments} & -0.047936 & -0.048911 & -0.053970
\\ \hline\hline
Mg-O increments & -0.019750 & -0.019798 & -0.019804 \\ 
O-O increments & -0.018516 & -0.017229  & -0.020154 \\ \hline
\multicolumn{1}{|c|} {two-body increments} & -0.038266 & -0.037027 & -0.039958
\\ \hline\hline
\multicolumn{1}{|c|} {three-body increments} & +0.000847 & +0.000818 &
+0.000862 \\ \hline\hline
\multicolumn{1}{|c|} {sum at 4.18 \AA} & -0.085355 &  -0.085120 & -0.093066 \\
\hline\hline
sum at 4.21 \AA  & -0.085233 & -0.084909 & -0.092983 \\
\hline \hline
\end{tabular}
\end{minipage}
\end{center}
\end{table}

\newpage
\begin{table}
\begin{center}
\begin{minipage}{5in}
\caption{\label{99}Local increments (in a.u.) for CaO
 at a lattice constant of 4.864 \AA }
\vspace{5mm}
\begin{tabular}{|c|rrr|}
\hline\hline
 & \multicolumn{1}{|c}{ACPF} & 
 \multicolumn{1}{c}{CCSD} & \multicolumn{1}{c|}{CCSD(T)}\\
\hline\hline
$\rm {Ca} \rightarrow \rm {Ca}^{2+}$ & +0.047046  &  +0.047359 &  
+0.050921 \\
$\rm {O}  \rightarrow \rm {O^{2-}}$ & -0.097060 & -0.098043 & -0.103456 \\
\hline
\multicolumn{1}{|c|} {one-body increments} & -0.050014 & -0.050684 & -0.052535
\\ \hline\hline
Ca-O increments & -0.037258 & -0.035258 & -0.040126 \\ 
O-O increments & -0.007527 & -0.006984  & -0.008263 \\ 
Ca-Ca increments  & -0.001038 & -0.001059 & -0.001179 \\ \hline
\multicolumn{1}{|c|} {two-body increments} & -0.045823 & -0.043301 & -0.049568
\\ \hline\hline
\multicolumn{1}{|c|} {three-body increments} & +0.000632 & +0.001662 &
+0.001484 \\ \hline\hline
\multicolumn{1}{|c|} {sum at 4.864 \AA} 
& -0.095205 & -0.092323 & -0.100619 \\ \hline
\hline
\multicolumn{1}{|c|} {sum at 4.81 \AA} & -0.096342 & -0.093445 & -0.101910 \\
\hline \hline
\end{tabular}
\end{minipage}
\end{center}
\end{table}

\newpage
\begin{table}
\begin{center}
\begin{minipage}{5in}
\caption{\label{88}Lattice constants of MgO and CaO (in \AA ). }
\vspace{5mm}
\begin{tabular}{|c|c|ccc|c|c|}
\hline\hline
System & RHF & ACPF  & CCSD & CCSD(T) & DFT & exp.  \\ \hline
MgO  & 4.191  \cite{Catti}
  & 4.181  & 4.173  & 4.184  & 4.11 ... 4.27  \cite{CausaZupan} 
& 4.2072  \cite{LandoltGitter} \\
 & & & & & 4.101 ... 4.105  \cite{McCarthy} & \\ \hline
CaO & 4.864  \cite{MackrodtCaO} & 4.809    & 4.805  & 4.801  
& 4.73 ... 4.85 
\cite{CausaZupan} & 4.8032  \cite{LandoltGitter} \\ 
\hline\hline
\end{tabular}
\end{minipage}
\end{center}
\end{table}

\clearpage
\newpage
\begin{figure}
\caption{Charge density of embedded O$^{2-}$ 
}
\end{figure}
\begin{figure}
\caption{Van der Waals-like decay of the two-body O-O increments in
CaO}
\end{figure}
\begin{figure}
\caption{Sum of local increments for MgO}
\end{figure}
\begin{figure}
\caption{Sum of local  increments for CaO}
\end{figure}
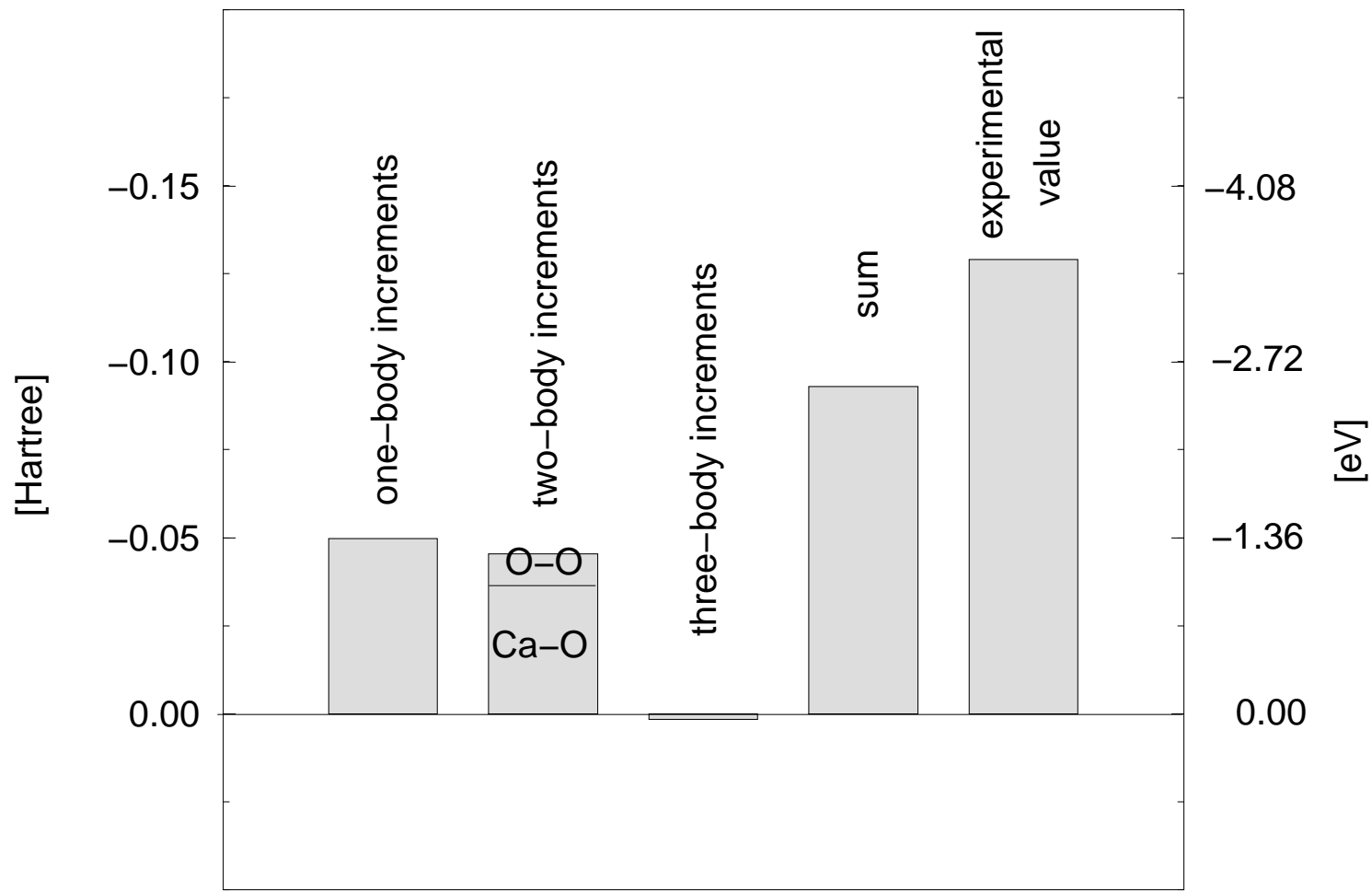
\end{document}